\documentstyle[prl,aps,epsf,multicol]{revtex}
\begin{document}
\draft

\def\i{\imath\,}
\def\ih{\frac{\imath}{2}\,}
\def\undertext#1{\vtop{\hbox{#1}\kern 1pt \hrule}}
\def\ra{\rightarrow}
\def\lfa{\leftarrow}
\def\ua{\uparrow}
\def\da{\downarrow}
\def\Ra{\Rightarrow}
\def\lra{\longrightarrow}
\def\ler{\leftrightarrow}
\def\lrb#1{\left(#1\right)}
\def\O#1{O\left(#1\right)}
\def\EV#1{\left\langle#1\right\rangle}
\def\tr{\hbox{tr}\,}
\def\trb#1{\tr\lrb{#1}}
\def\dd#1{\frac{d}{d#1}}
\def\dbyd#1#2{\frac{d#1}{d#2}}
\def\pp#1{\frac{\partial}{\partial#1}}
\def\pbyp#1#2{\frac{\partial#1}{\partial#2}} 
\def\pd#1{\partial_{#1}}
\def\br{\\ \nonumber & &}
\def\brr{\right. \\ \nonumber & &\left.}
\def\inv#1{\frac{1}{#1}}
\def\be{\begin{equation}}
\def\ee{\end{equation}}
\def\bea{\begin{eqnarray}}
\def\eea{\end{eqnarray}}
\def\ct#1{\cite{#1}}
\def\rf#1{(\ref{#1})}
\def\EXP#1{\exp\left(#1\right)} 
\def\INT#1#2{\int_{#1}^{#2}} 
\def\LHS{left-hand side }
\def\RHS{right-hand side }
\def\COM#1#2{\left\lbrack #1\,,\,#2\right\rbrack}
\def\AC#1#2{\left\lbrace #1\,,\,#2\right\rbrace}

\title{Fractionalization, topological order, and cuprate superconductivity}
\author{T. Senthil and Matthew P.A. Fisher}
\address{
Institute for Theoretical Physics, University of California,
Santa Barbara, CA 93106--4030
}

\date{\today}
\maketitle

\begin{abstract}
This paper is concerned with the idea that the electron is fractionalized
in the cuprate high-$T_c$ materials. We show how the notion of 
topological order may be used to develop a precise theoretical characterization
of a fractionalized phase in spatial dimension higher than one.
Apart from the fractional particles into which the electron breaks apart,
there are non-trivial gapped topological excitations - dubbed ``visons".
A cylindrical sample that is fractionalized
exhibits two disconnected topological sectors 
depending on whether a vison is trapped in the ``hole" or not.
Indeed, ``vison expulsion" is to fractionalization what the Meissner
effect
(``flux expulsion") is to superconductivity.
This understanding enables us to address a number of 
conceptual issues that need to 
be confronted by any theory of the cuprates based on fractionalization ideas. 
We argue that whether or not the electron fractionalizes in the cuprates
is a sharp and well-posed question with a definite answer. We elaborate on 
our recent proposal for an experiment to unambiguously settle this issue.

\end{abstract}
\vspace{0.15cm}


\begin{multicols}{2}
\narrowtext
  
\section{Introduction}
\label{intro}
The cuprate high-$T_c$ materials are amongst the most complicated 
systems studied extensively in solid state physics. 
In addition to the high temperature superconductivity itself,
they display a wide variety of novel phenomena.
Perhaps the most puzzling
is the behaviour in the ``normal'' non-superconducting state 
above the transition
temperature which, being anything but normal,
is difficult to understand within Fermi 
liquid theory. The
superconductivity is obtained by doping ``parent'' compounds that 
are Mott insulators -  rendered insulating by  
strong electron-electron interactions. These parent compounds also display 
Neel antiferromagnetism.  A number of other interesting phases
and broken symmetries are also often observed,
including charge and spin ordering into stripes. 
In addition, some regions of the phase diagram are very sensitive  
to the presence of disorder - particularly at low doping and low temperature. 
Indeed, even a casual glance at the phase diagram is sufficient
to realize the richness of phenomena displayed by these materials. 

It is hoped by many that underlying this remarkably 
complex behaviour, might lie a simple explanation
which will give insight into the mechanism of
the superconductivity.
The challenge is to identify any  
key {\it qualitative} features of the system
which can be sharply characterized and detected experimentally.
In this paper, we pursue an elegant and simple explanation of the superconductivity
and other properties that is based on the idea that the electron is 
splintered apart ({\em i.e} fractionalized)
in these materials.  The genesis of this idea can be traced
back to the original RVB theories\cite{PWA,KRS,LN}, but recent 
theoretical work\cite{RSSpN,Wen1,NLII,z2long,u1z2}
has lead to a unified theoretical framework for electron fractionalization
above one spatial dimension (most readily expressed
in terms of a $Z_2$ gauge theory\cite{z2long,u1z2}).  Remarkably,
this points to a
novel route
to superconductivity which dispenses entirely with
the notion of electron pairing.
Quite generally, to obtain superconductivity in a many-body system it is necessary
to condense a charged particle. In an electronic system
the naive route would be to 
condense the electron, but this is of course
not possible as the electron is a fermion. 
The BCS solution was to argue that a weak  
attractive interaction between the electrons (or more precisely
between Landau quasiparticles) binds them into pairs,
which condense as a charge $2e$ boson.
But fractionalization describes an altogether different
{\it route} to superconductivity,
within which the electron splinters into two
pieces, one carrying the Fermi statistics
(and spin) - a direct condensation of the
remaining charge $e$ boson leads directly to superconductivity.
Remarkably, although the fractionalization {\it route} to
superconductivity is so very different from that in BCS theory,
the resulting superconducting phase itself has identical qualitative
properties\cite{z2long,z2short}.  
Furthermore, the fractionalization idea provides appealing 
explanations of several of the unusual ``normal'' state phenomena, 
most notably the photoemission spectra.

In this paper, we show how a precise meaning may be given to the 
statement that the electron is fractionalized. 
Based on this, we argue that whether or not the electron fractionalizes
in the cuprates is a 
sharp theoretical question that is independent of all kinds of unavoidable 
material complications. Further, we 
show how this sharp theoretical question may 
be answered unambiguously by experiments. {\em The idea that the electron
is fractionalized thus provides a non-pairing route to superconductivity
which is directly 
testable}.

We begin by developing a precise theoretical characterization of a 
phase in which the electron is fractionalized. 
As anticipated in Ref. \cite{Wen1}, this is through the notion of
``topological order'' - a concept that has been elucidated clearly by Wen 
and coworkers\cite{Wen2} 
in the context of the quantum Hall effect. This enables us to address 
a number of conceptual issues that need to be confronted by any theory of 
the cuprates based on fractionalization ideas.
The crucial property of the fractionalized phase is the existence of excitations
which are fractions of the electron. While various such phases with different 
fractionalization patterns are theoretically possible\cite{pwv}, the phase that is of the most 
interest in the context of the cuprates is one in which the 
electron breaks into a charged boson and a neutral spin carrying fermion.
An equally crucial property of the fractionalized phase is the emergence of a 
gapped topological excitation - dubbed the vison\cite{z2long}. 
A pair of visons can 
annihilate each other, so that they carry only a $Z_2$ (topological) quantum number. 
The existence of this topological excitation is conceptually very 
important to the ``fractionalization'' route to superconductivity.
The superconductor obtained by condensing the {\it charge}
of an electron (once having shed it's Fermi statistics)
is in the same phase as the one obtained by condensing 
Cooper pairs of electrons. In particular, despite
the condensation of a charge $e$ boson,
flux quantization is in units of 
$hc/2e$  - 
this surprising result\cite{SNL,z2long} requires 
the presence of the topological
vison excitations in the fractionalized phase.
Indeed, the visons bind to an 
$hc/2e$ unit of electromagnetic flux once the system becomes superconducting.  

Any complete theory of the cuprates must necessarily pay attention to their
layered quasi-two dimensional structure.  Motivated by this, we consider 
the possible fractionalized phases in such a geometry. 
Interestingly, two qualitatively distinct kinds of fractionalized phases
are possible. In one, the system behaves as a full three dimensional solid, 
and the fractions into which the electron decays can freely propagate in all
three directions. In the other, the different layers decouple from each other  
- the fractions of the electron can now propagate freely within each layer
but cannot do so in the direction perpendicular to the layers. 
A number of experiments suggest that
this decoupled quasi-$2d$ fractionalized phase is the one more likely 
relevant to the cuprates.

Another important issue is the fate of the 
fractionalization at finite temperature\cite{z2long}. 
One normally associates 
fractionalization with a property of the spectrum of the 
system`s Hamiltonian - it is therefore not a priori clear 
whether it has any meaning at finite temperature. However,
having characterized the fractionalized
phase by it's {\it topological order} (rather than by it's spectrum)
we are able to address this issue. 
For the quasi-$2d$ fractionalized phase, the topological order 
in fact does not survive at finite temperature,
so that a {\it sharp} distinction between fractionalized and unfractionalized 
phases is only possible at zero temperature.  Nevertheless, at low temperature 
above the fractionalized phase, the system is ``almost"
topologically 
ordered. In the cuprates, we have suggested\cite{z2short} 
that the crossover towards the $T = 0$ fractionalization
occurs at a temperature comparable to the pseudogap temperature. 
As we will see, this may be exploited to probe the hidden zero temperature
order in the system. (For the fully three dimensional fractionalized
phase, on the other hand, the topological order survives up to a finite 
non-zero temperature\cite{z2long}). 
 
If fractionalization occurs at all in the cuprates, it is most likely
in the underdoped regime.
This might appear to raise
serious problems for the fractionalization idea, since 
it is precisely in the heavily underdoped region at low temperature that 
a variety of conventional broken symmetry states (Neel magnetism,
or charge and spin stripes) are observed. 
Furthermore, this region tends to be very sensitive to disorder effects. 
We argue that this is a non-issue. Theoretically, the topological
order that characterizes fractionalization can happily co-exist 
with Neel magnetism\cite{note0}, or stripes, or other broken symmetry states. 
Moreover, it is
unaffected by disorder. Thus, the presence of a conventional broken 
symmetry tells us nothing about whether or not the system 
is fractionalized at zero temperature. {\em If the electron is indeed fractionalized 
in the underdoped cuprates, the conventional ordered states seen in that region
are complications that distract from the 
hidden zero temperature topological order that is ultimately
responsible for the superconductivity.}

Historically, theoretical attempts to access fractionalized
phases above one dimension have focused on 
``quantum disordering" various states with
conventional well-understood 
broken symmetries, most frequently antiferromagnets and superconductors.
This has led to a misconception that fractionalization
{\it requires} the close proximity to a ``parent" conventional
broken symmetry state.
This, however, is both problematic and
incorrect.
Clearly there can be ``quantum
disordered" magnets or superconductors which
are {\it not} fractionalized.  Moreover, 
ordered phases which are fractionalized are certainly
possible, at least in principle.
As we emphasize in this paper, the correct way to 
characterize any fractionalized phase  
is by specifying it's {\it topological order}.
However, the fractionalized phase does often contain in it the seed of 
broken symmetry, particularly in electronic systems.
For example,
once the electron charge (or spin) has been liberated from it's
Fermi statistics, a direct condensation leads naturally to superconductivity
(or magnetism).
But note, here the broken symmetry {\it emerges} from the 
fractionalization - the latter being the higher energy
phenomenon. 
For instance, if fractionalization occurs at all in the
cuprates, the energy scale is presumably comparable
to the pseudo-gap temperature\cite{z2short} - and the superconductivity is an
emergent low energy phenomenon.
Thus it is  more correct to view the
fractionalized phase as the ``parent" phase to
the broken symmetry state - rather than the other way around.

While the underdoped cuprates are possibly fractionalized, the 
empirical evidence seems to suggest that when heavily overdoped
they are not. As we have detailed earlier\cite{z2short}, 
the quantum confinement
transition where the fractions of the electron get glued back together
might well account for the properties in the region between the underdoped 
and overdoped regimes. A complete theory of this novel quantum 
phase transition is 
unfortunately unavailable at present - we instead will briefly discuss 
some much simpler quantum confinement transitions.   

Most importantly, the theoretical understanding of fractionalization
developed in this paper enables us to describe
an experimental setup which should enable
a direct detection of the topological order.
As we shall see, the hallmark of fractionalization is the
expulsion of visons - analogous to the Meissner effect
being the hallmark of superconductivity.  We describe a way
to prepare and detect a vison in the hole of a cylindrical sample.
If the ``normal state" of the underdoped cuprates is
fractionalized, 
and hence topologically ordered, the trapped vison will
be unable to escape, and can be detected at a later time.
This signature of fractionalization in the ``normal state",
is directly analogous to fluxoid trapping
in a superconductor. Some of the results of this paper, mainly
the proposal for the experiment described above, were briefly
presented in a recent short paper\cite{toexp}.  

In the rest of the paper, 
we elaborate on the ideas and results described above. 
The theoretical formulation we use to describe fractionalization is a 
$Z_2$ gauge theory. While this is mathematically and physically 
closely related to several other formulations, it has several advantages. 
It works directly with the physical excitations in the fractionalized
phase.  Moreover, the topological order characterizing the fractionalized 
phases is most simply discussed in the $Z_2$ gauge theory framework. 
It also has the advantage that it generalizes readily to a variety 
of relevant situations,
such as layered systems or a system with broken
spin rotation invariance.

\section{Fractionalization and topological order}
\label{fto}
\subsection{Review of $Z_2$ formulation}
\label{z2rev}
In our recent work\cite{z2long} we demonstrated that a general class of
strongly interacting electron models could be recast in the form  
of a $Z_2$ gauge theory, which then enabled us to provide a 
reliable discussion of issues of electron fractionalization.
In particular, we demonstrated the possibility of obtaining fractionalized 
phases in two or higher spatial dimensions. We begin with a quick review 
of this formulation.  

The action for the 
$Z_2$ gauge theory is 
\bea
\label{IGA}
S & = & S_{c} + S_{s} + S_K + S_{B},  \\
S_{c} & = & - t_c \sum_{\langle ij \rangle} \sigma_{ij} ( b^*_ib_j + c.c.) ,\\
S_s &=& -\sum_{\langle ij\rangle} \sigma_{ij}
(t^s_{ij} \bar{f}_{i\alpha} f_{j\alpha} + t^{\Delta}_{ij} f_{i\ua}f_{j\da} + c.c ) - 
\sum_i \bar{f}_{i\alpha} f_{i\alpha} \\
S_K &=& -K \sum_{\Box} \prod_{\Box} \sigma_{ij}   .
\eea
Here, $b^\dagger_i$ creates a spinless, charge $e$ bosonic
excitation - the chargon - and $f^\dagger_i$ creates
the spinon, a fermion carrying spin $1/2$ but no charge.
When created together, these two excitations comprise the electron.
The field $\sigma_{ij}$ is a gauge field that lives
on the links of the space-time lattice
(taken as cubic when in $2+1$-dimensions), and takes on
two possible values:  $\sigma_{ij}=\pm 1$.  
The kinetic term for the gauge field, $S_K$,
is expressed in terms of plaquette products.  Here, $S_B$ is a Berry's 
phase\cite{z2long}
term which depends on the doping $x$.

At a formal level, the action above reformulates a system of interacting 
charge $e$, spin $1/2$ electrons
as a system of spinless, charge $e$ bosons (the chargons) and neutral,
spin $1/2$ fermions (the spinons) both of which are minimally 
coupled to a fluctuating
$Z_2$ gauge field. The physical content of any gauge field is in 
it's vortex excitations
that carry the gauge flux. We are therefore led to consider
vortices in the $Z_2$ gauge field - dubbed the ``vison''.
Specifically, consider the product of the gauge field $\sigma$ around 
an elementary plaquette, which can take on two values, plus or minus one.
When this product is negative, a vison excitation is present
on that plaquette.  

We may therefore regard the action in Eqn. \ref{IGA} above as a 
reformulation of an 
interacting electron system as a theory of interacting 
chargons, spinons, and visons.
At this stage, this is essentially nothing more than a 
change of variables on the original electronic system. However this reformulation
is an extremely useful starting point to discuss phases of the system where the 
electron is fractionalized. 
Both the chargons and spinons carry a unit of $Z_2$ gauge charge while the vison
carries a unit of $Z_2$ gauge flux. Thus, upon encircling a vison, the chargon and spinon
each acquire a phase of $\pi$. This long range interaction has crucial implications for the 
physics.

There are two qualitatively different phases that are 
described by the $Z_2$ gauge theory action.
In one, the visons are gapped excitations.  
In such a phase, the electron splits
into two independent excitations - the chargons and the spinons.  
To see this simply, consider the limit when the vison gap is very large so that 
they may be safely ignored (ie. $K \rightarrow \infty$). 
Thus, when the visons are
absent, all the plaquette
products of the $Z_2$ gauge field equal plus one. One can therefore
put $\sigma_{ij}=1$ on every link.  In this case the chargon
and spinon can propagate {\it independently}, and the electron
is {\it fractionalized}.

The other qualitatively different kind of phase is obtained if the visons are condensed. 
The long range interaction between the visons and the chargons (or the spinons)
frustrates the motion of the latter. The result is that they are confined together
to form electrons (or other composite excitations made out of electrons). In such a phase,
the electron is not fractionalized. Further, once the vison is
condensed, it loses it's legitimacy as an excitation in the system.

Thus the really crucial property of the fractionalized phase is the presence of 
the gapped topological vison excitations. 
The full excitation spectrum in the fractionalized phase
decomposes into different topological sectors. The fractionalized phase is 
therefore characterized by the emergence of a topological quantum number 
which labels the spectrum of states. 
Topological excitations are also well-known to occur in states with a broken symmetry
- for instance, vortices in superconductors. However the topological excitations in 
the fractionalized phase occur despite the absence of any obvious broken symmetry.
 
Nevertheless, the fractionalized phase contains in it the seed of broken symmetry. 
Once the electron is splintered into the chargon and the spinon, it's electric 
charge is no longer tied to it's Fermi statistics. Instead, the charge
is now carried by the bosonic chargons. The chargons can now directly condense
leading to a superconducting state. Surprisingly, this superconductor is in
the same phase as that obtained by the condensation of Cooper pairs of electrons. 
In particular, the superconductor has flux quantization in units of $hc/2e$
despite it's description as a condensate of charge $e$ chargons. 
This remarkable feature is due to the presence of the topological excitations - the visons
- in the fractionalized phase. Indeed upon condensing the chargon to form the superconductor,
the vison also acquires $hc/2e$ of electromagnetic flux. 

In the rest of this section, we will develop a precise 
theoretical characterization of the fractionalized
phase using the notion of topological order.
  
\subsection{Topological order in the pure gauge theory}
\label{pgt}
We begin by considering the pure gauge theory in the absence of any matter coupling
({\em i.e} coupling to the chargons or the spinons). This is described by the action
\be
\label{pgta}
S_K = -K \sum_{\Box} \prod_{\Box} \sigma_{ij}  .
\ee
For concreteness, we specialize to a two-dimensional
spatial square lattice, plus one time dimension.
It will often also be convenient to consider the equivalent 
quantum Hamiltonian\cite{Kogut} 
in two spatial dimensions:
\be
\label{pgth}
H = -  K\sum_{\Box} \prod_{\Box} \sigma^z_{rr'} - h \sum_{<rr'>} \sigma^x_{rr'} .
\ee
Here $r,r'$ label the sites of the $2d$ square lattice and $\sigma^z_{rr'}, \sigma^x_{rr'}$ are Pauli matrices that live on the bonds
of the lattice. The first term involves products over spatial plaquettes only.

It is well-known\cite{Kogut} that this pure $Z_2$ gauge theory 
has two phases. For $K$ small,
there is a  phase where static test charges that couple to the 
gauge field are confined. For $K$ large, on the other hand, there is a different phase
where such test charges are allowed to be deconfined. This distinction may be
quantified by the behaviour of the ``Wilson loop'' 
correlator\cite{Kogut} - this decays 
exponentially with the area of the loop in the small $K$ phase, 
but only with the 
perimeter in the large $K$ phase.

A different, but equivalent, view of these two phases is in terms of the 
vison excitation {\em i.e}, the vortex of the $Z_2$ gauge field. In the perimeter law phase, 
the vison is a gapped excitation. 
In the area law phase, on the other hand, the vison is condensed. This 
can be understood very explicitly by means of a duality transformation\cite{Kogut,note1} to 
the global Ising model described by the Hamiltonian:
\be
\label{tfim}
H = -h\sum_{r r'} v^z_r v^z_{r'} - K\sum_r v^x_r .
\ee
This global Ising model is defined on the lattice
dual to the original square lattice. The $v^z_r, v^x_r$
are also Pauli matrices. The dual Ising spin $v^z_r$ has the physical interpretation
of being the vison creation operator\cite{Kogut,z2long}. 
For small $K$, the global Ising model is in
it's ordered state, and the visons are therefore condensed. 
For large $K$, on the other hand,
the global Ising model is in it's disordered phase, and the visons are gapped.

The two phases of the gauge theory Hamiltonian in Eqn. \ref{pgth} 
may be distinguished in yet another way - this is through the notion of 
``topological order''. Consider the gauge theory Hamiltonian on a 
manifold with a non-trivial topology. In the deconfined (large $K$) phase,
as we discuss at length below,
the ground state has a degeneracy (in the thermodynamic limit)
which depends on the topology of the manifold. In the confined 
phase on the other hand, 
there is a unique ground state independent of the topology of the manifold. 
This is a precise, and as we shall see, powerful distinction between the 
two phases. Such a distinction was originally pointed out for pure (non-abelian) 
gauge theories 
in pioneering work by 't Hooft\cite{thooft}. 

This topological characterization of the phases of the gauge theory can be traced to 
the existence of symmetry operations specific to the topology of the manifold.
These topological symmetries are preserved by the ground state in the confined phase. 
In the deconfined phase, these topological symmetries are spontaneously broken
- this immediately leads to the ground state degeneracy on non-trivial manifolds. 
Such a breaking of topological symmetries also characterizes the fractional
quantum Hall fluids, as expounded in some beautiful papers\cite{Wen2} of 
Wen and coworkers. 
Following the terminology used in that context, we will refer to the 
breaking of the topological symmetry as ``topological order''.  
 
\begin{figure}
\epsfxsize=3.3in
\centerline{\epsffile{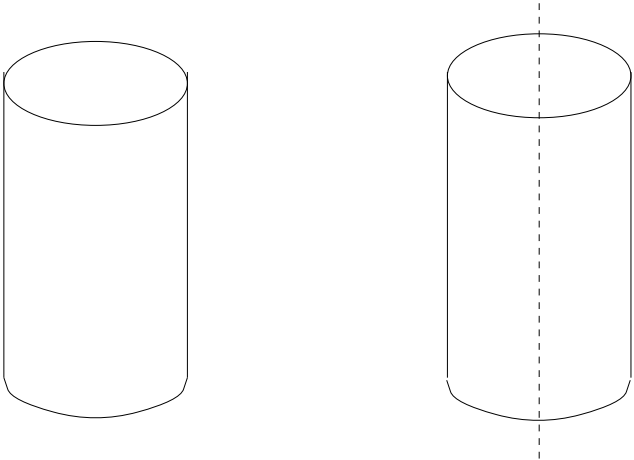}}
\vspace{0.15in}
\caption{The two degenerate states in a cylinder. The right one has a vison 
threading the ``hole''.}
\vspace{0.05in}
\label{cylv}
\end{figure}

\begin{figure}
\epsfxsize=1.5in
\centerline{\epsffile{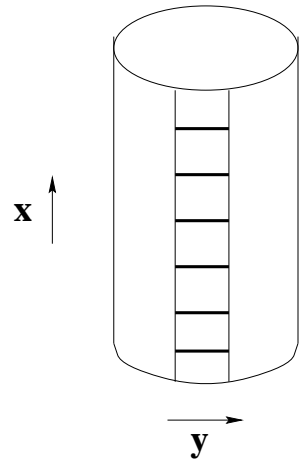}}
\vspace{0.15in}
\caption{Changing the sign of $\sigma^z$ on all the dark bonds adds
(or removes) a vison from the hole of the cylinder.}
\vspace{0.05in}
\label{cyl}
\end{figure}

To fix these ideas, consider a cylindrical geometry. In the deconfined phase 
of the gauge theory, there are two degenerate ground states. They correspond
to whether or not a vison has ``threaded the hole of the 
cylinder" (See Fig. \ref{cylv}).  Deep within the
deconfined phase, with $K \rightarrow \infty$, the
two corresponding gauge field configurations are very simple.
Setting all of the link fields, $\sigma^z =1$, is clearly a ground state
in this limit, and corresponds to the absence of threaded vison, since
the flux of the $Z_2$ gauge field through any curve $C$ that encircles
the cylinder,
\be
\label{flux}
\Phi[C] = \Pi_C \sigma^z_L
\ee
equals unity. (Here $L$ labels the bonds that belong to $C$.)
The ground state {\it with} a threaded vison can be obtained,
for example,
by changing the sign of $\sigma^z$ on a column of horizontal bonds
that runs the length of the cylinder (see Fig. \ref{cyl})  
- in this state $\Phi[C]= -1$.
Similarly reasoning implies that on a torus,
there are four degenerate ground states corresponding to the vison threading or not
threading each of the two holes. In what follows, we will analyse the cylinder 
in several ways to get a deep understanding of this phenomenon. 

\begin{figure}
\epsfxsize=3.3in
\centerline{\epsffile{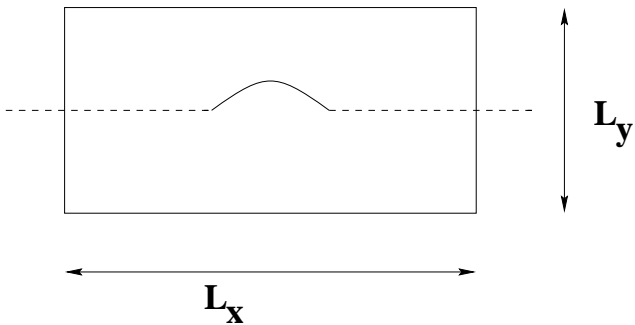}}
\vspace{0.15in}
\caption{Vison tunneling out of the cylinder. The dashed line  
represents the $Z_2$ flux line inside the hole. The points of penetration are
where the line becomes solid. It is assumed that there are periodic
boundary conditions along the $y$-direction.}
\vspace{0.05in}
\label{vtunn}
\end{figure}

Assume that a vison is initially trapped in the hole of the cylinder 
when the gauge theory is in it's deconfining phase. We take the 
$x$ axis of space to be along the length of the cylinder, and the 
$y$ axis to be along the circumference. Let the length of the 
cylinder be $L_x$ and it's circumference $L_y$.
For the vison to tunnel out, the $Z_2$ flux
tube must penetrate the cylinder in at least two places (in general some even
number), and these two points of penetration must move apart (see Fig. \ref{vtunn}) till
they drop out of the edge of the system. As there is a finite energy cost 
for the vison to penetrate the sample in the deconfined phase, the amplitude 
for this process will be exponentially small in $L_x$. Thus the 
vison tunneling rate varies as $\Gamma \sim e^{-c L_x}$, which goes to zero 
as $L_x \ra \infty$. Thus, once trapped, a vison in the hole of
the cylinder lives forever (in the thermodynamic limit).

Consider the situation with finite $L_x$ and $L_y$. An 
operator that adds (or removes) a vison from the hole can be  
readily constructed as follows:
\be
\label{vhc}
{\cal P} = \Pi_x \sigma^x_{\vec r, \vec r + \hat y}
\ee
with $\vec r = (x, y_0)$. The operator ${\cal P}$ changes the sign of 
all the operators $\sigma^z_{\vec r, \vec r + \hat y}$ that 
live on the bonds (see Fig. \ref{cyl}) 
along the $y$-direction between some chosen $y$-slices $y_0$ and 
$y_0 + 1$.  Consider now the flux of the $Z_2$ gauge field through any curve $C$
that encircles the cylinder, as defined in Eqn. \ref{flux}:
Clearly, ${\cal P}$ changes the 
sign of this flux. Thus ${\cal P}$ is an operator that adds or removes a vison from the hole
of the cylinder.

It is straightforward to see that ${\cal P}$ commutes 
with the Hamiltonian Eqn. \ref{pgth}
of the gauge theory. Thus, it is a symmetry of the theory. 
Further, as it corresponds to 
the operation of adding a vison through the hole, it is a topological symmetry. 
Now consider the limit $L_x \ra \infty$. As argued earlier, 
in the deconfined phase a vison 
that is trapped in the hole 
stays there forever. Consider the ground state with a vison trapped in the hole.
Upon acting on this state with the operator ${\cal P}$, it becomes the ground state
in the sector with no vison trapped. Thus, the ground state is not invariant under the 
action of the operator ${\cal P}$. The topological symmetry has been broken spontaneously. 
Note that the ground states in the two sectors (with or without a vison) 
are guaranteed to have
exactly the same energy as ${\cal P}$ commutes with the Hamiltonian. Thus, 
the gauge theory in it's deconfining phase has two degenerate ground states 
on the cylinder.

Further insight into the ground state degeneracy and the broken topological symmetry
is obtained by the following considerations. Imagine changing the value of $K$ to  
some $K_b$  
along all plaquettes at some $y$-slice, say $y = y_0$. 
Assume that the gauge theory in 
the bulk is in it's deconfining phase ({\em i.e}, the bulk value of $K$ is very large). 
In the limit that $K$ in the bulk is $\infty$, there 
can be no gauge flux penetrating the bulk
of the system. We may then set $\sigma^z = 1$ for all 
bonds except those along the ``cut''. 
The remaining degrees of freedom live on the cut (the dark bonds in Fig. \ref{cyl}). 
The Hamiltonian describing them is
clearly just a one dimensional transverse field Ising model:
\be
\label{cutham}
H = -K_b\sum_x \sigma^z_{x} \sigma^z_{x+1} - h\sum_x \sigma^x_x ,
\ee
where $\sigma^z_x$ is the $Z_2$ gauge field on the bond at site $x$ along the cut. 
For small $K_b$, this Ising model is in it's disordered phase. The ground state is 
therefore unique. With increasing $K_b$ this 
edge global Ising model undergoes a 
phase transition to an ordered state with $\langle \sigma^z_x \rangle \neq 0$. 
The ground state is therefore two-fold degenerate. The two degenerate ground states
correspond precisely to whether or not a vison is trapped in the hole
of the cylinder. This can be seen in several ways - for instance by noting
that the operator ${\cal P}$
introduced above is precisely the global spin flip operator of the edge Ising model. 
Further, the domain walls in the ordered state of the edge Ising model 
correspond to plaquettes where a vison has penetrated the cylinder. In the 
ordered phase, such domain walls, and hence the visons, cost finite energy. 
In the disordered state, the domain walls have proliferated - this may be 
interpreted as a proliferation and condensation of visons along the edge. 

The phase transition discussed above is thus an edge confinement transition. 
The topology of the manifold in which the deconfined phase ``resides'' changes 
from a rectangle to a cylinder as the coupling $K_b$ is increased. 
We will discuss such topology-changing phase transitions further in Section \ref{tcpt}.

\begin{figure}
\epsfxsize=3.3in
\centerline{\epsffile{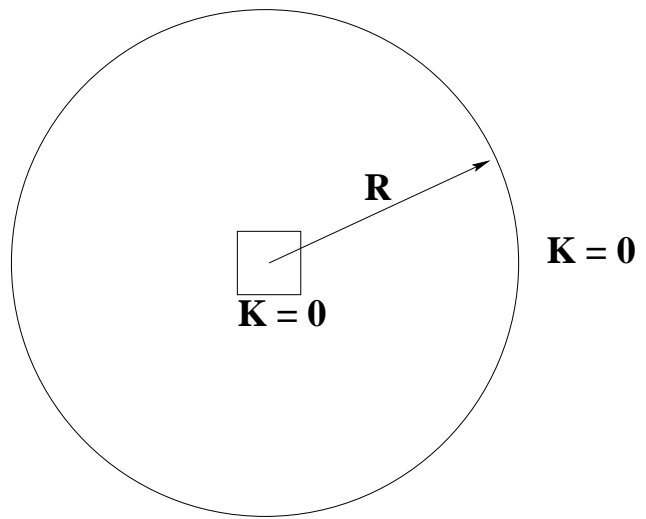}}
\vspace{0.15in}
\caption{The $Z_2$ gauge theory on an annulus with a hole at the center. This
corresponds to setting $K =0$ for the ``hole'' plaquette, and for all the 
plaquettes outside the disc radius.}
\vspace{0.05in}
\label{anngt}
\end{figure}

Yet another route to understanding the topological ground state degeneracy of the 
deconfined phase is to employ the duality transformation of the full
gauge theory to the global Ising model as discussed in the beginning
of this subsection. For this purpose, it is convenient to consider an 
annulus (see Fig. \ref{anngt}) which is topologically equivalent
to a cylinder.
This can be obtained from the 
gauge theory defined in infinite two-dimensional space by simply setting some of the plaquette strengths to zero. 
First, imagine setting $K = 0$  for a single plaquette in the center. This
creates a ``hole'' in the system. Similarly, at the outer boundary of the sample,
again set $K =0$ for all plaquettes. This captures the finiteness of the sample. 
For concreteness, we consider a circular disc of radius $R$.

Now we employ the duality transformation to get a representation of the system
as a global Ising model. The hole in the center of the sample goes over into 
a single site of the dual lattice. The restriction that $K =0$ at the hole then
implies that the transverse field at this site on the dual spin is exactly zero.
Similarly, at the outer boundary of the sample, $K =0$ implies that 
the transverse field on the dual Ising spins outside the disc radius is zero. 
This implies that these dual spins outside the disc radius are all
lined up together\cite{note2}. 

\begin{figure}
\epsfxsize=3.3in
\centerline{\epsffile{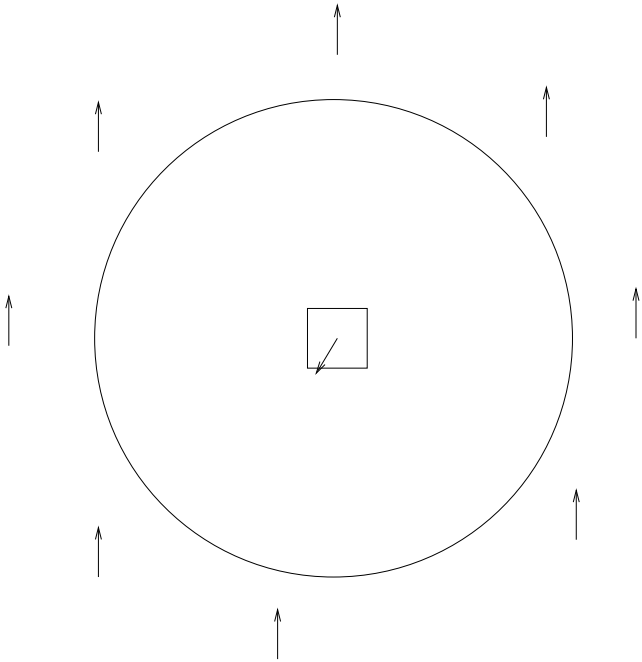}}
\vspace{0.15in}
\caption{The dual global Ising model on the annulus. The Ising spins 
outside the disc radius are frozen in the up direction. The Ising spin 
in the hole has no transverse field on it.}
\vspace{0.05in}
\label{annim}
\end{figure}

Before continuing, it is necessary to take note of one other subtle
feature of the duality transformation. Two states of the dual global Ising
model that only differ by an overall spin flip are not to be counted as 
two distinct states of the gauge theory (as may be seen
from, for instance, the treatment of the duality transformation in 
Ref. \cite{z2long}). This can be taken care of simply by 
fixing the direction of the frozen spins outside the disc radius to be, say,
up (see Fig. \ref{annim}).

We may now discuss the ground state degeneracies of the gauge theory 
using the dual global Ising model. 
First consider the confining phase of the gauge theory. This is the 
ferromagnetic phase of the dual global Ising model. The direction of the 
boundary spins fix the direction of the ferromagnetic ordering. Thus, 
all the spins in the interior including the one corresponding to the hole,
point in the up direction. There is no ground state degeneracy. 

Now consider the deconfining phase. This corresponds to the paramagnetic phase
of the dual global Ising model. The dual spin correlations decay exponentially.
Thus the ordering of the boundary spins has little influence on the ``hole'' spin 
at the center. The latter is therefore essentially free to point in any direction. 
This then corresponds to the expected two degenerate ground states. To make this 
more precise, consider modelling the bulk system by a continuum scalar
field theory with a Gaussian action,
\be
S = \int d\tau d^2x [ \left(\nabla \phi \right)^2 +    
  \left(\partial_{\tau} \phi \right)^2 + m^2 \phi^2 ]  .
\ee
This Gaussian theory is expected to correctly describe the physics
of the paramagnetic phase of the global Ising model. 
The coupling to the ``hole'' spin
is through an Ising exchange term,
\be
S_{hole} = \int d\tau v^z \phi(\vec 0, \tau) .
\ee
Here $v^z$ represents the ``hole'' spin and we have taken the 
location of the hole to be at the origin. Note that the ``hole''
spin has no dynamics - this is due to the absence of any transverse
field on that spin in the lattice model. 

The action above must be supplemented with a boundary condition 
arising from the fixed direction of the spins outside the disc radius. 
This is simply the condition that 
\be
\phi(\vec x, \tau) = \phi_0  ,
\ee
for $|\vec x| = R$ with $\phi_0$ a positive constant. 

As the field $\phi$ is massive, it can be safely 
integrated out to get an effective action for $v^z$. 
For result for large $R$ is simply,
\be
S_{eff} = 2\pi R \int d\tau \chi(R) v^z ,
\ee
where $\chi(R)$ is the static susceptibility of the 
Ising paramagnet. 
This is readily computed to be
\bea
\chi(R) & = & \int \frac{d^2q}{4\pi^2} \frac{e^{i \vec q . \vec R}}{q^2 + m^2} ,  \\
& = & \frac{1}{2\pi} K_0(mR)  , \\
& \sim & \sqrt{\frac{1}{8\pi mR}} e^{-mR} ,
\eea
where the last expression is valid for $R >> 1/m$. The effective action above for the 
``hole'' spin may be readily converted into an effective Hamiltonian,
\be
H_{eff} = \Gamma v^z ,
\ee
with $\Gamma \sim \phi_0 \sqrt{\frac{R}{m}} e^{-mR}$. 
Thus there are two low energy states with a splitting
$\sim \sqrt{R}e^{-mR} \ra 0$ as $R \ra \infty$. 

For large but finite $R$, the energy eigenstates are eigenstates of 
$v^z$. But if the system is prepared in one eigenstate of $v^x$, 
it takes a very long time (of order $\frac{1}{\Gamma}$) to tunnel to the other 
eigenstate. 

Physically, the operator $v^z$ adds or removes a vison 
from the hole of the annulus. The two eigenstates of $v^x$
correspond to a vison being either present or absent from the hole. 
In the confined phase, the ``hole'' spin is frozen in the up direction. 
The vison is therefore condensed in the hole, as it is in the rest of the sample. 
In the deconfined phase, a trapped vison (the ``hole'' spin in an eigenstate of 
$v^x$) stays in the hole for a time that diverges exponentially as 
the sample radius $R$ goes to infinity.

\subsection{Effect of matter fields}
\label{emf}
In the discussion above, we considered the phases of the pure $Z_2$ 
gauge theory and the topological distinction between them. We now put 
back the coupling to the chargons and spinon fields. In the presence of
such ``matter'' coupling, there continues to be a sharp distinction between 
the deconfined and confined phases. However, as is well-known\cite{Kogut}, 
the behaviour of the Wilson loop is no longer sufficient 
to distinguish the two phases once matter coupling is included. As we 
will see below, there is nevertheless a topological distinction between 
the two phases\cite{noteqcd}. 

Consider the properties of the system in a cylindrical geometry. 
Assume that the system is in 
it's deconfined phase. 
This implies that the vison is a gapped excitation. Consequently, 
a vison, once trapped in the hole of the cylinder, will stay there for a 
long time (of order $e^{cL_x}$) as argued previously.  
In the state with
no vison threading the hole of the cylinder, the chargons and spinons
are subject to periodic boundary conditions on encircling the cylinder. 
If, on the other hand, a single vison threads the cylinder, the chargons and spinons
are subject to antiperiodic boundary conditions. 
This difference in the boundary conditions leads to a {\em slight}
difference between the energies of the two states (with or without
a vison threading the hole). However, this energy difference {\em vanishes} 
in the thermodynamic limit. Thus, the ground state is two-fold degenerate
in the thermodynamic limit. 

To put some meat into these observations, we restrict attention to the 
ground states in the two topological sectors with or without a vison in the hole, and 
denote these as $| \uparrow \rangle$ and  
$| \downarrow \rangle$, respectively.
The Hamiltonian when projected to this 
subspace may be written,
\be
\label{tls}
H_{proj} = \Gamma \tau^x + h \tau^z ,
\ee
where $| \uparrow \rangle$ and  
$| \downarrow \rangle$ are the two eigenstates of the Pauli matrix $\tau^z$.
Clearly, the Pauli matrix  $\tau^x$ is the operator\cite{note3} that adds or removes a vison 
from the hole.
The first term therefore corresponds to the tunneling
of the vison, with tunneling rate $\Gamma \sim e^{-cL_x}$ as established
in the previous subsection.  The term proportional to $\tau^z$ comes from 
the difference in energy between periodic and antiperiodic boundary conditions
for the chargons and spinons. The dependence of the splitting $h$ on the dimensions
of the system is determined by the properties of the spectrum of the chargons 
and the spinons. In the insulating phases of interest, the chargon is always gapped. 
If the spinon is also gapped, then it is easily seen
that $h \sim e^{-\tilde{c} L_y}$. Note that this splitting vanishes exponentially 
in the cylinder {\em circumference} while the vison tunneling rate vanishes  
exponentially in the cylinder {\em length}. In a fractionalized phase with 
linearly dispersing gapless spinons (as happens in the nodal liquid or the $d-RVB$ state),
the splitting vanishes only as $h \sim \frac{L_x L_y}{L_y^3}$. The inverse dependence
on the linear system size may be guessed by scaling considerations: 
Indeed the low energy theory is simply a Dirac theory for the nodal spinons. 
This theory is critical with a dynamic critical exponent $z = 1$. 
Consequently, the energy $h$ vanishes inversely with the linear system size. 
This argument may also 
be verified by an explicit tedious computation\cite{dbdcalc} on a representative lattice model.  

The projected Hamiltonian has two eigenvalues
\be
E_{\pm} = \pm \sqrt{h^2 + \Gamma^2}
\ee
Clearly, the splitting between these two levels goes to zero in the thermodynamic limit
leading to two degenerate ground states. 

It is important to note that the term $h \tau^z$ which arises due to the presence of
matter coupling explicitly breaks the topological symmetry discussed in the 
previous subsection. Indeed, in the restricted space above, the topological 
symmetry is implemented by the operator $\tau^x$. This no longer commutes 
with the Hamiltonian when matter fields are present. However, the commutator
goes to zero as the system size goes to infinity. Thus, 
we may view the operation
of threading a vison through the hole as becoming a good (topological) symmetry 
in the thermodynamic limit, which is then spontaneously broken. 
While this is, in principle, a correct point of view, it is not entirely
satisfying. 

The more crucial point to note is that there are two 
distinct topological sectors in the cylinder (with or without a vison) 
in the deconfined phase even in the presence of the chargons and spinons. 
This is simply the statement that a trapped vison stays there 
forever in the deconfined phase. In the confined phases, on the other hand, 
a trapped vison is absorbed by the vison condensate, and is very quickly lost.
Therefore, there is no topological quantum number labelling the states (other than 
those associated with any conventional broken symmetry that may be present).

It is also useful to consider the system in an annulus geometry with a 
finite-sized hole at the center. Here again, in the deconfined phase, a 
vison that is trapped in the hole stays there forever (when the outer radius 
of the annulus goes to infinity). However, now there is a finite energy 
difference between the states with and without a trapped vison due to the 
change in the boundary conditions on the chargons and spinons
upon encircling the hole. Thus the inability
of the trapped vison to escape is really the hallmark of the fractionalized phase. 
The experiment proposed in Ref. \cite{toexp} that we elaborate on in
Section \ref{dto} probes precisely this property.  
 
Before concluding this section, we note that a ground state degeneracy 
of four on a torus was suggested\cite{ReCh} to exist for certain states described by 
specific RVB wavefunctions. The same result was shown\cite{RSSpN} to obtain in the 
phases of frustrated spin models that show fractionalization. In these fractionalized
phases, there are neutral spin-$1/2$ excitations that have {\em Bose} statistics. 
Evidently, in this case, 
fractionalization has liberated the spin from the Fermi statistics of the electron. 
Despite the similarity in the ground state degeneracy, the topological order that 
characterizes this phase is {\em distinct} from that of the phases of primary interest
in this paper. This may be seen by using {\em gedanken} flux-trapping experiments 
of the kind discussed in Ref. \cite{toexp} (see also Ref. \cite{pwv}).

\section{Layered systems}
\label{Ls}

Among the many unusual properties of the cuprate materials is the
stark difference between the in-plane and c-axis transport.
Both at optimally doped and in the slightly underdoped regime,
the normal state often exhibits ``metallic" in plane transport - 
with the resistance dropping upon cooling - which co-exists with insulating
c-axis transport.  As emphasized by Anderson\cite{PWAbook}, 
this behavior is difficult to reconcile with
a conventional Fermi liquid picture of the normal state,
particularly in the low temperature limit (accessed by suppressing
the superconductivity with strong field) where in-plane coherence 
of Landau quasiparticles would be expected to eventually
lead to coherenct c-axis motion as well.
Motivated by this puzzling behavior, we consider in this section
issues of fractionalization in an anisotropic layered system.
Quite strikingly, we argue that two distinct
fractionalized phases are possible - one 
which exhibits deconfinement of spinons and chargons in all three
spatial directions, and another quasi-two-dimensional fractionalized
phase in which the spinons and chargons are deconfined within
each layer but cannot propagate coherently between layers.
In this section we restrict attention to zero temperature,
turning briefly to the effects of thermal fluctuations
in Section \ref{fte}.

For simplicity, we will follow the strategy adopted in Section \ref{fto},
and initially consider the pure $Z_2$ gauge theory - appropriate
to the layered geometry - before incorporating
the spinons and chargons into the theory.
To this end, consider the Hamiltonian for a $Z_2$ gauge theory
defined on a $3d$ cubic lattice appropriate to
an anisotropic layered system:
\begin{equation}
{\cal H} = - K_{xy} \sum_{P_{xy}} \prod_{P_{xy}} \sigma_{r r^\prime}^z - K_{\perp}
\sum_{P_{\mu z}} \prod_{P_{\mu z}} \sigma_{r r^\prime}^z 
- h \sum_{\langle r r^\prime \rangle} \sigma_{r r^\prime}^x     .
\end{equation}
Here the first term is a sum over all plaquettes in the $x-y$ plane
(normals along the z-axis) and the second term is a sum over all other
plaquettes  (normals lying in the $x-y$ plane, with
$\mu = x,y$).  For simplicity we have taken the transverse field 
strength to be the same for all
links of the $3d$ spatial lattice.

\begin{figure}
\epsfxsize=3.3in
\centerline{\epsffile{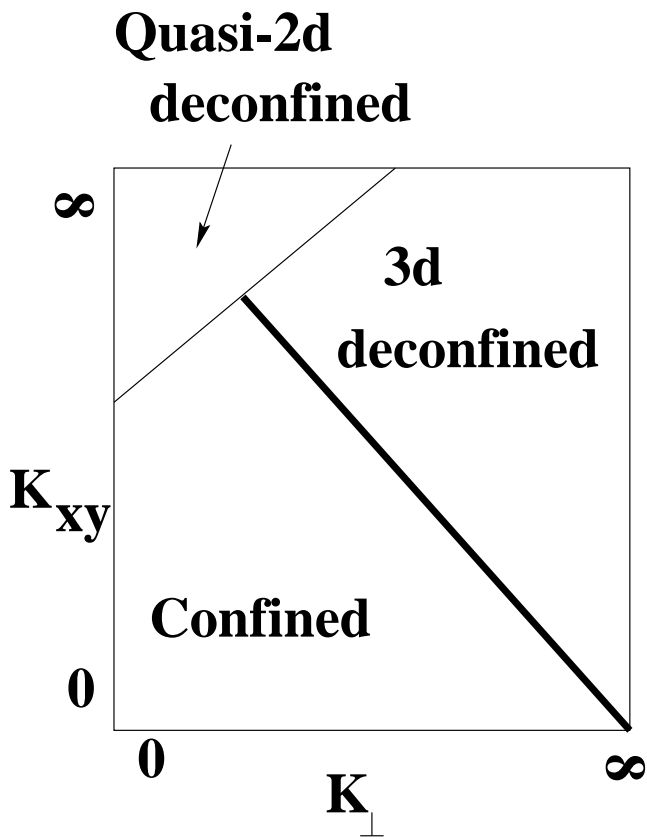}}
\vspace{0.15in}
\caption{Schematic phase diagram of the $Z_2$ gauge theory in a layered
geometry. The solid line is a first order phase transition.}
\vspace{0.05in}
\label{lyrphs}
\end{figure}

As defined, this Hamiltonian depends on just two dimensionless parameters,
$K_{xy}$ and $K_{\perp}$ measured in units of the transverse field $h$.
The ground state phase diagram in this two-dimensional space of couplings can be
readily inferred by considering various simplifying limits.
For example, when $K_{\perp} =0$ the trace over $\sigma^x$ on the interlayer 
links can be trivially performed, and the model reduces to a set
of decoupled $2+1$ dimensional gauge theories, one for each layer.
Then, each layer has two phases - a confined phase for small $K_{xy}$
and a deconfined phase for large $K_{xy}$, as depicted schematically in Fig. \ref{lyrphs}.
Away from the intervening transition, one expects  
the distinction between
these two phases to survive for small non-zero $K_{\perp}$.  In both phases,
vison loops proliferate {\it between} the layers, so that the spinons
and chargons which carry the $Z_2$ charge cannot move coherently
along the $c-$axis.  For small $K_{xy}$ the vison loops can also freely 
penetrate
the layers, so that spinons and chargons are confined in all
spatial directions.  But the phase with large $K_{xy}$ (and small $K_{\perp}$)
is most unsual:  Since the interlayer vison loops are expelled
from the layers, the in-plane motion of the spinons and chargons is coherent,
but they are nevertheless confined along the $c-$axis.  

To see how 
this unusual quasi two-dimensional deconfined phase survives
with small non-zero $K_{\perp}$, we consider other limiting regimes
of the phase diagram.
Along the diagonal with $K_{\perp}=K_{xy}\equiv K$, the $Z_2$ gauge theory Hamiltonian
describes an isotropic three-dimensional situation whose phase diagram is well
understood - there is a first order transition at $K=K_c$ of order one separating
the fully confined phase at small $K$ from a three-dimensional deconfined phase.
In the deconfined phase all large vison loops are expelled, and the spinons
and chargons can propagate coherently in all three directions.
  
Now consider the
limit of infinitely large $K_{xy}$.
When $K_{xy}=\infty$, the $Z_2$ flux is forbidden from penetrating
the $xy$ plaquettes (thereby restricting the vison loops
to lie between successive layers).  It is therefore 
possible to choose a gauge
in which $\sigma^z = 1$ on all links lying in the $xy$ plane.
The system then decouples into a set of $2d$ sub-systems, which live between
adjacent layers. Consider specifically the Hamiltonian for a single such $2d$ 
subsystem,
which depends on the gauge fields living on the interlayer links which can
be labelled
conveniently by a $2d$ square lattice of sites denoted $\bbox{r}$:
\begin{equation}
\label{tdil}
{\cal H}_{2d} = - K_{\perp} \sum_{\langle \bbox{r} \bbox{r}^\prime \rangle}
\sigma_{\bbox{r}}^z \sigma_{\bbox{r}^\prime}^z - h \sum_{\bbox{r}} \sigma_{\bbox{r}}^x  .
\end{equation} 
Notice that the plaquette product term has reduced to a near-neighbor
Ising coupling in this sub-system Hamiltonian.  Indeed, ${\cal H}_{2d}$ is
precisely a 2d transverse field quantum Ising model, which exhibits
two phases as the ratio $K_{\perp}/h$ is varied. The two phases
are separated by a $2+1$-dimensional Ising phase transition.
It is clear that the interlayer vison loops of the original
anisotropic gauge theory, are simply domain walls separating regions with
positive and negative Ising ordering, $\sigma^z = \pm 1$. In the
ferromagnetically ordered phase of the transverse field Ising model
with large $K_{\perp}$  the interfacial energy
is non-vanishing. It follows that large interlayer vison loops are 
excluded - this
is the $3d$ deconfined phase as depicted in Fig. \ref{lyrphs}.
But for small $K_{\perp}$ in the paramagnetic phase of the Ising model,
the interfacial energy vanishes.  In this case, the interlayer vison 
loops unbind and proliferate.
This is the anisotropic quasi-$2d$ deconfined phase
(discussed above
at large $K_{xy}$ and $K_{\perp} \rightarrow 0$). For large but 
finite $K_{xy}$ both deconfined phases will
continue to exist. 
Piecing together the above results, one arrives at the final phase diagram for
the anisotropic layered $Z_2$ gauge theory,
as drawn schematically in Fig. \ref{lyrphs}.

In passing we note that phases very similar to the decoupled layered  
phase discussed above have been considered in other contexts in the 
literature. For a $U(1)$ lattice gauge theory, precisely such a phase was argued
to exist when the spatial dimension of each layer is at 
least three in Ref. \cite{fu_niel}. 
In a different context, recent work\cite{dll} has examined the stability of ``decoupled 
Luttinger liquid'' phases in  quasi one dimensional systems. In the context of
cuprate physics, the possibility of such decoupling of the layers has been 
emphasized by Anderson and coworkers\cite{PWAbook}.

It is illuminating to briefly consider the topological ordering that characterizes
the three phases.  In the $3d$ deconfined phase, since the vison loops are fully expelled,
one expects a two-fold ground state degeneracy when periodic boundary conditions
are imposed in any one of the three spatial directions,
say along the $c$-axis (with open boundary
conditions along the other two directions).  As discussed for the $2d$ 
gauge theory
in Section \ref{fto} above, in the thermodynamic limit the two states correspond to
the presence or absence of a vison loop threading the "hole" in the cylinder.
When the transverse linear dimension $L$ is finite, there will be a small
splitting of order $exp(-cL^2)$, due to the tunnelling of an
interlayer vison loop.  More generally, in a $d-$dimensional deconfined phase,
one expects a tunnel splitting varying as $exp(-cL^{d-1})$.  As 
emphasized by Wen\cite{Wen1,Wen2},
the power in the exponent is particular to topological ordering,
and should be contrasted to the $exp(-cL^d)$ splitting between the two
states of a model with a local order parameter such as the $d-$ dimensional
transverse field quantum Ising model.  When periodic boundary conditions are taken
along all three spatial dimensions, the ground state degeneracy in the $3d$
deconfined phase is of course $2^3 = 8$.

In the confined phase of the $3d$ gauge theory the ground state is unique
independent of the boundary conditions indicative of an absence of any topological
ordering.  But the topological ordering
that characterizes the quasi-$2d$ deconfined phase is somewhat subtle.  With periodic
boundary conditions only along the $c-$axis the ground state is unique,
due to the proliferation and condensation of interlayer visons in this phase.
This can also be understood more formally as follows:   
Consider the operation 
$\sigma^z \rightarrow - \sigma^z$,
which changes the sign
of all the interlayer bonds between (any) two adjacent layers.
This is the precise equivalent for the $c-$direction of the operator
${\cal P}$ introduced in Eqn. \ref{vhc} of Section \ref{fto}, and
clearly changes the sign of the $Z_2$ flux enclosed
by any curve that encircles the cylinder. Thus
this operation adds or removes a vison from the hole of 
the cylinder. As before, it commutes with the full Hamiltonian. 
But notice that in the $K_{xy} \rightarrow \infty$
limit, this transformation is simply a global Ising spin flip
for the $2d$ interlayer Hamiltonian given in Eqn. \ref{tdil}.  In the quasi-$2d$
deconfined phase, the interlayer quantum Ising models are disordered.
This implies that the ground state is invariant under the operation of 
threading a vison through the hole of the cylinder, and is hence unique.

Next consider the topological order in the quasi-$2d$ deconfined
phase when  
periodic boundary conditions are imposed in the plane,
say just along the $y$ direction.  To understand the resulting ground state
degeneracy, it is simplest to first consider a model with {\it two}
layers only, which can be conveniently
visualized as two concentric cylinders with $y-$periodic boundary conditions 
around the cylinder.
Moreover, we specialize to the
$K_{xy} \rightarrow \infty$ limit which precludes
visons loops from penetrating either layer.  
One then expects that there should be $2^2=4$ low energy states 
which belong to topologically distinct sectors. 
These are distinguished by the presence or absence of a vison loop
threading through the bore of either concentric cylindrical shell.
To establish that these four states are in fact degenerate
(in the thermodynamic limit) first note that the symmetry operation
which adds a vison to both shells simultaneously (implemented
in each layer as in Eqn. \ref{vhc} of Section II), commutes with the 
two-layer gauge theory Hamiltonian. 
This implies that these four states
are in any case pairwise degenerate.  It remains to establish, though,
that the state with no visons has the same energy as the state
in which (only) one of the two cylindrical shells has a threading
vison.  To see this, note that the operation which threads
a vison through one layer only is equivalent (at $K_{xy} = \infty$)
to changing
from $y-$periodic to antiperiodic boundary conditions in the
interlayer $2d$ quantum Ising model of Eqn. \ref{tdil}.  Since this Ising
model is in it's disordered phase in the quasi-$2d$ deconfined
phase, the energy change will clearly be exponentially
small in the cylinder diameter ($exp(-cL_y)$).  This vanishes in the 
thermodynamic limit, thereby establising the degeneracy of all
four states.  For a layered system with $N$ layers,
the ground state degeneracy with $y-$periodic boundary conditions
in the quasi-$2d$ deconfined phase is simply $2^N$.

Upon inclusion of the spinon and chargon matter fields which carry
$Z_2$ charge, the nature of the topological ordering effects 
the inter and intra-layer confinement.  In the $3d$ deconfined phase
the chargons and spinons can propagate coherently in 
all three spatial directions. As before, with periodic boundary conditions,
they are sensitive to the presence or absence of 
visons trapped in the holes.  In the confined phase free spinons and chargons
cannot exist. But in the quasi-$2d$ deconfined phase although the
spinons and chargons can propagate coherently in-plane, they
are confined to live in one layer only.  The inter-layer $c-$axis transport
of the chargons and spinons is fully incoherent.
In this quasi-$2d$ phase it is possible to integrate out 
the gauge fields living on the vertical links (trivially so when
$K_{\perp} \rightarrow 0$). At second order in the ratio
of the interlayer spinon and chargon hopping amplitudes to
the transverse field, $h$, one generates
inter-layer electron and pair hopping terms as well as inter-layer 
magnetic exchange interactions.
As the chargons cannot propagate along the $c-$axis,
one would expect qualitatively
different inter and intra-layer charge transport at finite temperatures in this novel
quasi-$2d$ fractionalized phase.  We now turn to a brief discussion of finite
temperature effects.

\section{Finite temperature effects}
\label{fte}
In previous sections, we have discussed a precise theoretical
characterization of quantum phases (in $d \geq 2$)
where the electron is fractionalized. There is a sharp 
distinction between fractionalized and un-fractionalized
phases at zero temperature. Does this sharp distinction survive at
finite non-zero temperatures? One normally thinks of
fractionalization in terms of the spectrum of excitations
of the Hamiltonian describing the system. It is then not clear what
meaning it has at finite temperature. However, characterizing the 
fractionalized phase by it's topological order enables us to 
address this issue. 

We again start by considering the pure $Z_2$ gauge theory 
in two spatial dimensions. 
As with conventional broken symmetries, 
thermal fluctuations play an important role
in symmetry restoration when a topological symmetry is 
spontaneously broken.  Moreover, broken topological symmetries
are likewise less robust against thermal fluctations 
in low dimensions.   
Since the visons are (gapped) point-like excitations in the
topologically ordered $2d$ deconfined phase with 
a {\it finite} energy gap,
there will be a non-vanishing density of visons created thermally
at any non-zero temperature.  
This will immediately destroy the topological order.
The situation is loosley analogous to the quantum Ising model in
$1d$, which breaks the Ising spin-flip symmetry only
exactly at zero temperature.  The topological symmetry restoration due to
the thermally excited excited visons is perhaps easiest to understand
in the $2d$ annulus geometry.  There, at zero temperature the
two topologically ordered sectors correspond to the presence
or absence of a vison trapped in the hole of the annulus.
Clearly, at finite temperature a vison trapped in the hole of the annulus
can be thermally excited into the bulk, and can then leave the sample
at the outer edge of the annulus - this process interconnects
the two $T=0$ states with broken topological symmetry.

In the presence of chargon and spinons matter fields
the energy cost of a vison is still finite, so
quite generally the $2d$ topologically order will be 
destroyed at $T \ne 0$.  Nevertheless, as discussed in Ref. \cite{toexp} and 
Section \ref{dto},
by performing measurements at
``short" enough timescales it should be possible to detect the presence of
the $T=0$ topological order
at temperatures well below the vison gap.

Remarkably, the topological order in the deconfined phase
in three spatial dimensions survives thermal fluctuations intact\cite{z2long}.
Since the gapped vison excitations are {\it loops} in this case,
they are much more difficult to thermally excite.  Indeed,
the energy cost of a loop grows
linearly with it's length, $L$,
as does the entropy associated with the loop.  Thus, at low 
enough temperatures
the free energy tension of the loop will be positive,
effectively suppressing long unbound loops. 
Again, this reasoning remains valid in the 
presence of coupling to matter fields.  As the temperature is
raised eventually the vison loop entropy will dominate, and the
system will undergo a true finite temperature phase transition
at $T=T_c \ne 0$ that restores the topological symmetry.  
For $T < T_c$ in this $3d$ topologically ordered phase,
the free energy of two
$hc/2e$ magnetic monopole ``test" charges will grow linearly
with their separation - $hc/2e$ magnetic monopoles
are thus confined!  However, an {\it even} number of
magnetic monopoles - with flux an integer multiple of $hc/e$ - 
is not a source of vison flux and so costs only a finite energy.
Thus, it is only a $Z_2$ magnetic charge which is
confined in the $3d$ topologically ordered state.
This should be contrasted with the situation in a 
3d superconductor, which confines magnetic monopoles
with any magnetic charge - a $U(1)$ magnetic confinement.

Finally, we address the effects of thermal fluctuations
in the quasi-$2d$ deconfined phase that can occur
in a layered system (such as the cuprates).  Here, the topological order
is due to the suppression of vison loops penetrating
through the layers.  But
the energy cost for a vison loop to pass through a layer
is finite, and so will occur with non-vanishing density at
at any finite temperature.  Thus, strictly speaking,
quasi-$2d$ topological order in a layered system will
be destroyed at any non-zero temperature,
just as in the $2d$ case.

In Ref. \cite{toexp}, we described in detail an experimental signature
of this quasi-$2d$ topological order, which should allow
for it's detection if present in the underdoped cuprates.
We elaborate on this further in Section \ref{dto}. 
The presence of $T=0$ quasi-$2d$ topological order
shoud also lead to dramatic differences between 
the low temperature in-plane and $c-$axis transport.

\section{Coexistence with other broken symmetries}
\label{coex}
If fractionalization of the electron occurs at all in the cuprates, it 
does so in the underdoped portion of the phase diagram.
Furthermore, the fractionalized phase is presumably of the quasi-$2d$
kind discussed at length in Section \ref{Ls}. This implies that 
the associated topological order, strictly speaking, exists 
only at zero temperature. On the other hand, empirically, it is
precisely in the heavily underdoped region at low temperature that 
a variety of conventional broken symmetry states are observed. The undoped 
cuprates show Neel antiferromagnetism. At intermediate doping, charge and spin
stripe
instabilities have been reported. Furthermore this region is often also
thought to be disorder dominated. These observations raise the following 
conceptual questions: Can fractionalization coexist with conventional broken symmetry?
Is fractionalization possible in a disordered system? Armed with the 
precise theoretical characterization of the fractionalized phase expounded
in this paper, we now 
discuss the former question. The effect of disorder is considered in the following
section.

Once the electron has splintered into the chargons and the spinons, 
various kinds of charge ordering determined by the strong Coulomb interactions
between the chargons is certainly possible. Away from a doping level that 
is commensurate with the underlying lattice, such a charge-ordered insulating state
will break lattice translational and rotational symmetries. Thus, it is obvious
that fractionalization can coexist with charge order. 

A more interesting issue, first raised by Balents et.al.\cite{NLII}, is the 
possibility of coexistence of fractionalization and antiferromagnetism
or other kinds of magnetic order. In principle, this can be induced by 
interactions between the gapless spinons in the nodal liquid or d-RVB
state. If such a fractionalized antiferromagnet (dubbed $AF^*$) does exist,
what is it's precise distinction with the conventional Neel antiferromagnet
(dubbed $AF$)? Consider, in particular, the situation where the 
antiferromagnetic ordering wavevector connects two antipodal nodal
points of the spinons. Then, in the presence of Neel ordering, the spinons acquire
an energy gap. In this case, there would seem to be no distinction between 
$AF$ and $AF^*$ at low energies. Indeed, both phases would have gapless spin
wave excitations with a linear dispersion. 

The distinction is actually topological - the phase $AF^*$ has a topological order
(and the related vison excitations) that is not shared by the phase $AF$. 
This may again be seen by asking for the ground state degeneracy on, say, a torus of size
$L \times L$. (For simplicity, we specialize to two spatial dimensions).
Due to the long range Neel order, there will be the usual tower of states\cite{PWA_af}
scaling as
\be
\label{tower}
E_S = \frac{\lambda S (S + 1)}{L^2} ,
\ee
where $S$ is the total spin of the state, and $\lambda$ is a constant. 
These states should exist in both $AF$ and $AF^*$. But the phase $AF^*$ must have 
an additional four-fold degeneracy corresponding to trapping or not 
trapping a vison in each hole of the torus. Once a vison is trapped 
in a hole of the torus, it tunnels out at a rate $\Gamma \sim e^{-cL}$. 
The presence of a vison in the hole does not affect the magnons at any energy 
(as they are created by operators bilinear in the spinons), but it does affect
the boundary conditions of the gapped spinons. This results in a difference
$h \sim e^{-\tilde{c}L}$ between the energies of  states with and without a 
vison trapped in a hole. Thus, as explained in Section \ref{fto}, 
there are four 
states with a splitting that vanishes {\em exponentially} with $L$. 
This is to be contrasted with the tower of states above which approach 
zero as $1/L^2$. Furthermore, all these four states will have $S = 0$.

As noted above, the heavily underdoped cuprates 
exhibit several kinds of conventional broken symmetry - including the Neel ordering
at zero doping, and charge and spin stripes at finite doping. 
The discussion above shows that it is theoretically possible that 
the fractionalization and the associated topological order 
coexist with these coventional broken symmetries. This is 
conceptually very important - the fractionalization of the electron
provides a direct route to superconductivity that doesn't invoke ideas 
of pairing. If the heavily underdoped cuprates are fractionalized, 
then the Neel antiferromagnetism and the striping,
while interesting phenomena, 
are side issues not directly related to the origin of the superconductivity.

\section{Disorder}
\label{dirt}
One of the remarkable aspects of superconductivity
is the relative insensitivity of the Meissner effect 
to microscopic details, such as the symmetry of the
underlying crystal structure or the presence of impurities and defects.
Provided the superfluid density is non-vanishing,
expulsion of magnetic flux (and of vorticity) persists.
As we now discuss, the topological order that characterizes
a fractionalized phase is likewise insensitive to 
impurity scattering.  Since the essence of fractionalization
is the expulsion of topological visons, just as the essence
of superconductivity is the expulsion of vorticity, this 
insensitivity to dirt is perhaps not
surprising.

We focus our discussion on the deconfined phase in two spatial
dimensions.  To address the issue of the stability 
of topological order to dirt we consider
the pure $Z_2$ gauge theory in Eqn.~\ref{pgta}, since coupling in the
chargons and spinons will not change the essential energetics
of the visons.  In a spatially inhomogeneous system with impurities
present, the coupling constants $K$ and $h$ in the $Z_2$ gauge Hamiltonian
will vary randomly.  The dual global Ising model in Eqn.~\ref{tfim}
likewise becomes random - a
$2$-dimensional transverse field quantum Ising model with
quenched random bond strengths.  Upon inclusion
of a doping dependent Berry's phase term in the gauge theory, the Ising bond strengths can be negative, which leads to frustration.  
With one electron per site
the dual global Ising model is actually fully frustrated, and with randomness
present will effectively be a $2-$dimensional quantum Ising spin-glass.   
But recall that the deconfined phase actually corresponds
to the {\it paramagnetic} phase of the dual global Ising model - the
phase in which the visons (the Ising spins) are gapped out
rather than condensed.  The Ising paramagnetic phase is clearly
stable in the presence of random bonds.  Frustration from the negative
Ising bonds will likewise not destroy
the paramagnet, and might in fact actually
enhance it's stability.  As mentioned above,
inclusion of matter couplings will not modify this.
We thereby establish the important conclusion:  Topological order
that characterizes electron fractionalization in two dimensions
is robust and survives in the presence of impurity scattering.
The $3d$ and quasi-$2d$ deconfined phases considered in earlier
sections are likewise stable to dirt.  This fact is critical when one
considers searching for signatures of topological order in the
very underdoped cuprates, which are often riddled with defects
and charge inhomogeneities (eg. stripes).

\section{Topology-changing phase transitions}
\label{tcpt}
In recent work we have suggested that the unusual normal state properties
of the {\it optimally} doped cuprates might possibly be due to a
direct quantum phase transition between a $d-$wave superconductor
and a Fermi liquid.  As discussed in Ref. \cite{z2short}, this strong coupling
phase transition should be thought of as a "quantum confinement critical point".
On the deconfined side of the transition the electron fractionalizes
into chargons and spinons, and a subsequent condensation of
the bosonic
chargon leads
to superconductivity.  At the quantum critical point the chargons and spinons
become confined together recovering the electron, and one enters a Fermi liquid
phase.  Unfortunately, the critical properties of this most interesting
confinement transition are very difficult to access.  In this section we 
revisit the two much simpler quantum confinement transitions mentioned in Section's
II and III, and briefly address their critical properties.
Since topological order present in the fractionalized phase disappears
upon undergoing a confinement transition, these 
can be thought of as ``topology changing" phase transitions.

\subsection{Two-dimensions}
\label{twod}
Perhaps the simplest possible topology changing phase transition is the one explored
briefly in Section II.  For a $2d$ cylindrical sample 
in a deconfined phase with a ``cut" of weakened
bonds running parallel to the axis of the cylinder, there are two phases:
(i) A topologically ordered phase with a two-fold degenerate ground state
when the bonds along the cut are strong;
(ii) A phase with a unique ground state and no topological order
when the bonds are weak.  In the latter phase, the chargons and spinons
cannot propagate coherently across the cut, and are thus
deconfined on a topologically trivial manifold (the 2d plane),
in contrast to the former case where the chargons and spinons
can be taken coherently around the cylinder.

As detailed in Section II, for the pure $Z_2$ gauge theory
which is deep within the deconfined phase, the effective $1d$ theory
across the cut is simply the $1d$ transverse field quantum Ising model.
The quantum confinement transition corresponds to the ferromagnetic
to paramagnetic transition in the Ising model, and is in the universailty
class of the $D=1+1$-dimensional classical Ising model.

In the presence of {\it gapped} chargon and spinon matter fields one
does not expect the universality class of this transition to be modified.
But more interesting behavior becomes possible in a ``nodal liquid"
(or d-wave RVB) phase in which the deconfined spinons are
gapless at the four nodal points.  In this case one can readily write
down an effective field theory that should describe the critical
properties of this boundary confinement transition, by coupling
the spin of the $1+1$-dimensional quantum Ising model to the spinon hopping
across the cut.  Schematically, the effective action should take the form:
\begin{equation}
S = \int dx dy d\tau [ {\cal L}_{spinon} + {\cal L}_{Ising} + {\cal L}_{int} ]  ,
\end{equation}
with a $2+1$ Dirac form for the spinons\cite{note4},
\begin{equation}
{\cal L}_{spinon} = \Theta(y) \psi^\dagger_1 \partial \psi_1
+ \Theta(-y) \psi^\dagger_2 \partial \psi_2   ,
\end{equation}
where $\psi_1$ and $\psi_2$ are nodal spinors on the two sides of the boundary,
and
\begin{equation}
{\cal L}_{Ising} = \delta(y) [ (\partial_\mu \phi)^2 + r\phi^2 + u\phi^4 ]  ,
\end{equation}
is a soft-spin $1+1$ quantum Ising model ($\mu = x,\tau$).
The (schematic) form of the boundary coupling is:
\begin{equation}
{\cal L}_{int} = t_b \delta(y) \phi [\psi^\dagger_1 \psi_2 + c.c. ] .
\end{equation}  
When $r < 0$ the Ising field picks up a non-zero expectation value,
$\langle \phi \rangle \ne 0$, and the spinons can propagate coherently
across the cut.  For $r >0$ the Ising model is disordered, and one can
integrate out the $\phi$ field, generating a spin exchange interaction
across the boundary - the spinons are confined on either side of the
boundary, however.  The boundary confinement transition
occurs at $r=0$ (within mean field theory).

The critical properties can be accessed by considering a simple
renormalization group transformation which rescales
both spatial coordinates and time by the same factor.
When $t_b =0$, the theory decouples into a (critical) massless
$2+1$-dimensional free Dirac theory and a critical $1+1$-dimensional Ising model.  The relevancy
of a small interaction across the cut can then be deduced in terms
of the scaling dimension of the Dirac field ($\Delta_\psi = 1$)
and the Ising field ($\Delta_\phi = 1/8$):
\begin{equation}
\partial t_b /\partial \ell = (2-2\Delta_\psi - \Delta_\phi ) t_b  .
\end{equation}
Thus, the spinon hopping amplitude is actually an {\it irrelevant} perturbation,
scaling to zero with eigenvalue $-1/8$.  Being irrelevant, the transport
of spinons across the cut right at the confinement transition can be
deduced by working perturbatively in $t_b$.

\subsection{Inter-layer confinement transition}
\label{ilct}
The situation is somewhat more interesting when gapless spinons are present
at the confinement transition separating the $3d$ deconfined phase from
the quasi-$2d$ deconfined phase in an anisotropic layered situation
(like the cuprates).  The simplest situation to consider is that
of a layered system with {\it two} layers only.  To access the critical
properties it is sufficient to consider
the limit that $K_{xy} = \infty$, so that
visons cannot penetrate through either layer.  The remaining $Z_2$
gauge degrees
of freedom live on the interlayer bonds, and are described by the
$2+1$-dimensional quantum Ising model, Eqn. \ref{tdil}.  The Ising spin
is coupled to the interlayer spinon hopping.  An effective field theory
can be easily written down, taking a very similar form to above,
except with,
\begin{equation}
{\cal L}_{spinon} = \psi^\dagger_1 \partial \psi_1
+  \psi^\dagger_2 \partial \psi_2   ,
\end{equation}
where now $\psi_1$ and $\psi_2$ are nodal spinors in the two layers,
and
\begin{equation}
{\cal L}_{Ising} =  (\partial_\mu \phi)^2 + r\phi^2 + u\phi^4  ,
\end{equation}
is a $2+1$-dimensional quantum Ising model ($\mu = x,y,\tau$).
The interaction term is (schematically)
\begin{equation}
{\cal L}_{int} = t_b \phi [\psi^\dagger_1 \psi_2 + c.c. ]  .
\end{equation}
Once again, as above, one can consider a simple RG transformation
which leaves the massless $2+1$ Dirac and critical $2+1$ Ising theories
invariant.  Since the boundary tunnelling interaction is now
over the $2d$ spatial plane, the eigenvalue of $t_b$ is modified as,
\begin{equation}
\partial t_b /\partial \ell = (3-2\Delta_\psi - \Delta_\phi ) t_b  ,
\end{equation}
with $\Delta_\psi =1$ as above, but now
$\Delta_\phi \approx 0.52$ is
the scaling dimension 
of the spin field for the $2+1$ critical Ising theory.
In this case the interlayer interaction is quite strongly relevant,
and one will crossover to a strongly interacting critical theory.
One might be able to access this critical point by generalizing the
Dirac and Ising
theories to general $D=d+1$ dimensions, and expanding
around a Gaussian theory perturbatively
in $D=4-\epsilon$ space-time dimensions.

\section{Detection of topological order}
\label{dto}
In previous sections, we have discussed how a precise {\em theoretical}
characterization of fractionalized phases may be obtained through
the concept of topological order. In a recent paper\cite{toexp}, we 
proposed an experiment that will directly probe this topological order.  
This enables a precise {\em experimental} characterization of fractionalized
phases. In this section, we will discuss this experiment at length, 
providing more details than available in Ref. \cite{toexp}
and considering extensions. 

\begin{figure}
\epsfxsize=3.3in
\centerline{\epsffile{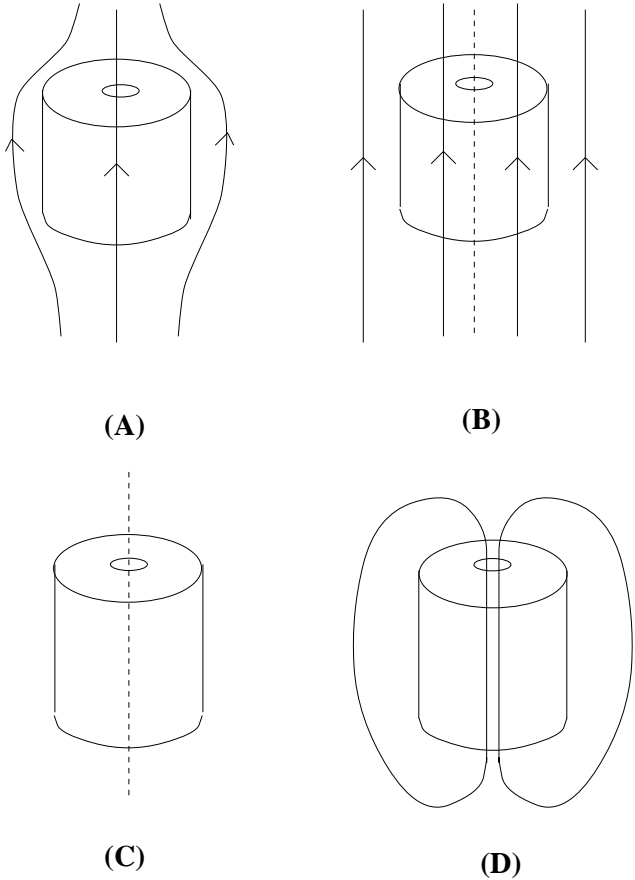}}
\vspace{0.30in}
\caption{The experiment to detect the topological order. 
The sample is superconducting in Figure A and Figure D and is 
``normal'' in Figure B and Figure C. The electromagnetic flux is shown as 
a solid line. In A, a flux of $hc/2e$ is trapped in the hole. On moving to B,
the electromagnetic flux penetrates, but the vison (shown as dashed line) is still
trapped. In C, the sample is in zero external magnetic field, but still has a trapped vison.
On moving back to the superconductor in D, a spontaneous flux of $hc/2e$
appears - it's direction is arbitrary.}
\vspace{0.5in}
\label{vd}
\end{figure}

The crucial property of the fractionalized phase is the inability 
of a trapped vison to escape from the cylinder. The effect described in 
Ref. \cite{toexp} is a direct probe of this property and involves the following 
sequence of events (see Fig. \ref{vd}): 

(a) Start with an underdoped sample
in a cylindrical geometry, with the axis of the cylinder
perpendicular to the layers.
In the presence of a magnetic field,
cool into the superconducting phase
such that exactly 
one $hc/2e$ magnetic flux quantum is trapped in the hole of the cylinder. 

(b) Heat the sample to above $T_c$.  

(c) Now turn off the magnetic field.  

(d) Cool the sample back down below $T_c$. 
 
An alternate experiment is to again repeat the sequence of events (a) to (d), 
but now work at a fixed very low temperature
and move from the superconductor into 
the (underdoped) insulator, and back, by adiabatically 
tuning some parameter.

In the non-superconducting state at the end of step (b), 
the magnetic flux penetrates
into the sample\cite{note5}. If,  however, this state is 
topologically ordered, then a $Z_2$ flux, {\em i.e} a vison, remains trapped. 
(Recall that the vison is bound to the $hc/2e$ vortex inside the 
superconductor.)  On turning off the magnetic field in step (c), time reversal invariance 
is achieved. What we have done is to prepare the sample in the 
non-superconducting state with a vison trapped in the hole of the cylinder.
This imposes antiperiodic boundary conditions on the chargons and spinons.
On moving back into the superconductor in step (d)
where the chargon condenses, the vison cannot exist by
itself and must nucleate an $hc/2e$ unit of magnetic flux. This {\em breaks}
the time reversal invariance achieved in step (c). The direction of the 
spontaneous flux is independent of that of the initial flux.

This spontaneous appearance of a magnetic flux is a direct consequence of 
the inability of the trapped vison to escape in a topologically ordered 
phase. We have, in effect, used the superconducting state to 
prepare and detect the vison\cite{notsc*}. 
In particular, if the non-superconducting state does not 
have the topological order, 
then there will be no spontaneous flux. 

In the cuprates, the fractionalization is 
presumably of the quasi-$2d$ kind discussed in Section \ref{Ls}. 
Thus, strictly
speaking, the topological order exists only at zero temperature. 
In this case, 
if the experiment is performed by tuning some parameter 
to reversibly move across the 
superconductor-insulator phase boundary at very low 
temperature, a spontaneous flux
is certainly expected. This is, however, much more challenging than 
cycling with temperature. What will be the outcome 
of the experiment done by 
varying the temperature? At a low but non-zero temperature, the trapped 
vison will eventually escape out of the sample in 
some time $t_v$. A spontaneous flux will be seen if the 
time scale for the experiment
is smaller than $t_v$.  

Decay of the 
trapped vison requires thermal activation across the vison gap in the bulk
of the sample. Thus 
\be
t_v \sim t_0 e^{\frac{E_0}{k_B T}} ,
\ee
where $T$ is the temperature, and $E_0$ is the vison gap. The 
prefactor $t_0$ is a microscopic time scale that depends much more weakly
on temperature. Thus $t_v$ increases strongly with decreasing temperature.
 
How big is the vison gap? 
A {\em lower} bound on this gap may be obtained from the results of 
ARPES studies of the underdoped cuprates. One of the most striking 
features of these experiments is the absence of a quasiparticle peak in the 
non-superconducting state. This is indeed as expected at low temperatures 
below the vison gap in a fractionalized phase. The ARPES intensity 
continues to be broad all the way up to the pseudogap
temperature $T^*$. This suggests that the vison gap is 
{\em at least} as big as $T^*$. 
In earlier work\cite{z2short}, we have suggested that the 
observed pseudogap crossover in the underdoped cuprates 
actually occurs at the scale of the vison gap, 
{\em i.e} $E_0 \sim k_B T^*$. 
  
A reliable estimate of the time $t_v$ is difficult
in view of the exponential sensitivity to the ratio of the vison gap
to the temperature. But the discussion above does suggest that 
$t_v$ can be enhanced enormously by enhancing the ratio $T^*/T_c$
and performing the experiment at temperatures close to $T_c$. 
A promising candidate material would therefore be 
$Bi_2Sr_2CaCu_2O_{8+x}$ ($ Bi2212$). In the heavily underdoped regime
when $T_c \sim 10 K$, a value of $T^* \sim 300 K$ has been reported\cite{jcc}.

A number of other equally robust predictions can be made for small
modifications of the experiment, as also discussed in Ref. \cite{toexp}. 
In particular, if the experiment is done with an initial flux of 
$\frac{nhc}{2e}$, a spontaneous flux of $hc/2e$ will be observed 
for $n$ odd at the end of the experiment, while no spontaneous flux will
be observed for $n$ even. This even/odd effect may be useful to 
rule out other mundane 
explanations of the effect, such as the presence of unknown stray 
magnetic fields in the sample at the end of step (c). 
A further observation is that the effect will not be observed 
if the axis of the cylinder is {\em parallel} to the layers. This 
is because, with quasi-$2d$ fractionalization, vison loops are 
condensed in the region between the layers. A vison that is 
initially trapped parallel to the layers will then be quickly absorbed 
by this vison condensate and escape. 

\subsection{Two holes and quantum tunneling of visons}
It is also extremely interesting to consider the situation where there are 
{\em two} holes drilled into the sample separated by a distance $l$ much smaller
than the sample radius $R$. To begin with, we specialize
to a strictly two dimensional system. Imagine starting in the superconducting state
with a single $hc/2e$ flux quantum trapped in one of the two
holes. Upon moving to the non-superconducting state either by heating 
or by other means, the magnetic flux penetrates into the sample. But again, 
if this non-superconducting state is fractionalized, the vison will be 
expelled from the bulk of the sample. However, in this case, the vison can 
tunnel back and forth between the two holes. 
Consider this experiment done at
zero temperature by moving reversibly between the superconducting and non-superconducting
phases. Then the tunneling of the vison from one hole to the other is entirely 
quantum mechanical. {\em It is therefore not possible even in principle to 
predict with certainty which hole the vison will be in after a given amount of time}. 
The best that can be done is to predict the probability of the vison being in any given hole.
Now, on reentering the 
superconducting state, the vison again acquires an $hc/2e$ unit of electromagnetic flux.
However, the resulting $hc/2e$ vortex can no longer tunnel so
readily between the two holes.
Now a measurement of the flux trapped will see a $hc/2e$ unit of flux in one
or the other hole.

Thus the two-hole experiment offers an opportunity to probe  
quantum tunneling phenomena at a macroscopic scale. The superconductor 
is used to prepare and detect the presence of a vison. Once the 
non-superconducting state is prepared in a state with a vison in one hole, 
it evolves quantum mechanically into a state which is a linear superposition
of the two states with the vison being in either hole. Moving back into the
superconductor nucleates $hc/2e$ flux which can be used to detect the 
presence of a vison. The relation of the observed probability for the 
flux being in either hole to the original vison wavefunction (in the 
non-superconducting state) depends on the details of the dynamics of the system,
and we will not discuss it here.

In the more complicated situation with several layers, the visons in each layer
can tunnel independently between the two holes. At the end of the experiment, one frozen-in
$hc/2e$ flux line will still be observed. This will pass through one of the 
two holes in each layer. The detailed shape of the flux line is an intriguing
question that we leave open for the present.

\section{Conclusions}
\label{concl}

In this paper, we have addressed a number of conceptual 
issues related to the possibility of electron fractionalization 
in spatial dimensins higher than one. Before concluding, we summarize some 
of the main results.   

The precise theoretical characterization of a fractionalized phase is 
through the notion of topological order. Apart from the fractional 
particles into which the electron breaks apart, there are 
non-trivial gapped topological excitations - the visons. The full excitation
spectrum therefore decomposes into different topological sectors. 
If a vison is initially trapped in the ``hole'' of a cylindrical sample 
that is fractionalized, it stays there forever.  

Motivated by the strongly anisotropic behaviour of the cuprates
in the non-superconducting states, we considered the possible fractionalized 
phases in a layered geometry. Interestingly, there are two kinds 
of fractionalized phases. In one, the system behaves like a full three 
dimensional solid with the chargons and spinons being able to freely propagate
in all three directions. In the other phase, the different layers decouple
from each other. The chargons and spinons are deconfined in each layer, 
but are confined in the direction perpendicular to the layers. 
It is this quasi-$2d$ deconfined phase
that is quite possibly relevant to the cuprates. 

We also considered the effect of a non-zero temperature 
on the topological order. For the quasi-$2d$ deconfined phase, 
the topological order does not, strictly speaking, 
survive at finite temperature. However, at temperature scales much smaller
than the zero temperature vison gap, it is ``almost'' topologically ordered. 
In the cuprates, we have suggested\cite{z2short} that the vison gap
sets the scale for the pseudogap crossover. 

We argued that the fractionalization could coexist with various
conventional broken symmetries, and even in the presence of disorder. 
Again, the notion of topological order gives a precise characterization
of ordered fractionalized phases (such as the phase $AF^*$) which 
distinguishes them from the corresponding ordered phases 
without the fractionalization. 

We also briefly discussed some toy examples of quantum confinement transitions.
The motivation was that precisely such a transition might possibly 
control the finite temperature properties of the cuprates in the region
between the under and overdoped regimes.   

One of the main points made in this paper is that 
the electron fractionalization idea provides a simple but directly testable
explanation of the superconductivity in the cuprates. 
We now briefly review the basis for this statement. 

(i) Fractionalization of the electron liberates it's charge from 
it's Fermi statistics. The resulting charged boson can then directly condense
leading to superconductivity. This is an alternative to the pairing 
route to superconductivity. 

(ii) Despite the alternate mechanism, the resulting 
superconductor is in the same phase as one obtained by condensing Cooper
pairs of electrons. That this is true may appear surprising 
given that what is condensing is a charge $e$ boson (rather than a charge 
$2e$ one). In particular, the flux quantization is in units of $hc/2e$. 
This remarkable feat is made possible by the presence of gapped 
topological excitations - the visons - in the fractionalized phase.
Thus the existence of these excitations is crucial for the fractionalization
route to superconductivity.  

(iii) The experiment we propose directly detects the stability 
of a trapped vison in the ``normal'' state of the cuprates. 

In view of the above, we believe that it should be possible to 
definitively establish or rule out the fractionalization explanation
of cuprate superconductivity. 

We are grateful to Leon Balents and Doug Scalapino for many
stimulating discussions.
We would like to thank John Kirtley for
his insights as to the feasibility of the proposed
vison trapping experiment.
This research was generously supported by the NSF 
under Grants DMR-97-04005,
DMR95-28578
and PHY94-07194.

\end{multicols}

\begin{references}
\bibitem{note0}This possibility was first considered in Ref. \cite{NLII}.
\bibitem{note1}Some readers may find the treatment in 
Appendix C of Ref. \cite{z2long} more illuminating.
\bibitem{noteqcd} This is true more generally than the specific
$Z_2$ gauge theory considered here. In particular, for non-abelian gauge theories  
in $D = 3+1$ with massive quark matter the ground state degeneracy in a three dimensional 
box with periodic boundary conditions is presumably 
different in the deconfined and confined phases. 
This precise distinction, even with matter fields, does not seem to be widely appreciated.
\bibitem{note3} This is simply the projection of the operator ${\cal P}$ to 
the restricted Hilbert space of the two ground states.
\bibitem{notsc*} If the superconducting phase adjacent to the fractionalized
phase is itself an exotic ``fractionalized superconductor" 
(the $SC^*$ phase
introduced in Ref. \cite{z2long}), then it cannot be used to detect
a trapped vison.
\end{references}
\end{document}